\documentclass{iau}
\usepackage{graphicx}

\title[Radio Galaxies, AGN, and SMBHs] 
{Radio Evidence for AGN Activity: \\ Relativistic Jets as Tracers of SMBHs}

\author[K. I. Kellermann]   
{K. I. Kellermann}

\affiliation{National Radio Astronomy Observatory, \\ 520 Edgemont Rd., Charlottesville, VA, USA \\ email: {kkellerm@nrao.edu} }

\pubyear{2015}
\volume{312}  
\pagerange{119--126}
\jname{Star Clusters and Black Holes in Galaxies and Across Cosmic Time}
\editors{Fukun Liu ed.}
\begin{document}

\maketitle

\begin{abstract}

Although the radio emission from most quasars appears to be associated with star forming activity in the host galaxy, about ten percent of optically selected quasars have very luminous relativistic jets apparently powered by a SMBH which is located at the base of the jet.  When these jets are pointed close to the line of sight their apparent luminosity is enhanced by Doppler boosting and appears highly variable.   High resolution radio interferometry shows directly the outflow of relativistic plasma jets from the SMBH.  Apparent transverse velocities in these so-called ``blazars'' are typically about 7c but reach as much as 50c indicating true velocities within one percent of the speed of light.  The jets appear to be collimated and accelerated in regions as much as a hundred parsecs downstream from the SMBH.  Measurements made with Earth to space interferometers indicate apparent brightness temperatures of $\sim 10^{14}$ K or more.  This is well in excess of the limits imposed by inverse Compton cooling.  The modest Doppler factors deduced from the observed ejection speeds appear to be inadequate to explain the high observed brightness temperatures in terms of relativistic boosting.

\end{abstract}

\firstsection 
\section{ Why Radio? }
Historically the first speculations about the existence of active galactic nuclei (AGN) and super massive black holes (SMBHs) came from the huge energy requirements implied by the discovery of distant powerful radio galaxies and quasars.  Today, radio observations remain crucial to understanding the role of SMBHs in astrophysics.  Only at radio wavelengths is it possible to image the region immediately surrounding the SMBH central engine and the relativistic jets which apparently originate with the SMBH.  Typical resolution obtained with Very Long Baseline Interferometer (VLBI) observations at centimeter wavelengths is of the order of 0.001 arccsecond (1 milliarcsec).  Thus, for nearby sources such as those located in the Virgo cluster, a linear resolution of  1 milliarcsec corresponds to only about 0.1 parsec or about 100 Schwartzchild radii for the SMBH located in the nucleus of M87.  For $z \sim 1$ the resolution for Earth-based VLBI is $\sim $ 10 kpc.  However, using an Earth to space interferometry at $\sim$ 1 cm or Earth-based systems at millimeter wavelengths, the resolution is improved by more than another order of magnitude.  Finally, we note that at radio wavelengths, there is no obscuration, even close to the SMBH.

However, not all radio sources are due to AGN; and not all AGN and SMBHs are radio sources.  Nearly all observed extragalactic radio emission is probably due to synchrotron radiation from ultra relativistic electron with energies $\sim$ 10 GeV moving in weak magnetic fields with $B\sim 10^{-5}$ to $10^{-4}$ Gauss.  The high energy electrons are thought to be accelerated in one of two ways; either by a central engine associated with accretion onto a SMBH in elliptical  galaxies or in quasars, or by supernovae following massive star formation (starbursts) in the  nucleus of early type galaxies.  Unfortunately, both processes are often referred to as AGN, and this has led to considerable confusion in the literature. 

Above $\sim 1$ mJy, the radio source number-flux density relation is dominated by sources driven by SMBHs. These more powerful sources are characterized by extended radio lobes, and by highly beamed relativistic jets extending from a few parsecs to hundreds of kiloparsecs from the SMBH. At microJy levels there is an increasing contribution from star formation related activity rather than from a SMBH.  However, there is an uncomfortably large spread in the observed microJy source count even among different observers using the same instrument, the VLA, in the same field.  Most likely these discrepancies are the result of systematic errors in the reported flux densities due to uncertainties in corrections for the effects of resolution \cite{CCF12}.

A recent complexity comes from the ARCADE 2 balloon measurement of a 3 GHz sky brightness of $54 \pm 6$ K which is about 5 sigma above that expected from known radio sources, suggesting a possible population of previously unrecognized weak sub microJansky sources \cite[(Fixen et al. 2011)]{F11}.  However, deep 3 GHz VLA observations showed no evidence for any source population greater than about 30 nanoJy.  Any population of weaker sources that could produce the excess sky brightness would need to have a sky  density greater than 6 x $10^4$ per ster, or 60 times greater than the density of the faintest (mag 29) galaxies in the Hubble Ultra Deep Field \cite[(Condon et al. 2012)]{CCF12}.  Thus if the excess background temperature observed by ARCADE is real and due to discrete sources, these sources cannot be associated with any known galaxy population. It will be important to verify that the ARCADE 2 results were not contaminated by unrecognized Galactic radio emission. 

\

Radio emission due to AGN can usually be distinguished from that due to star formation in a variety of ways.

$\bullet$ \ {\bf Morphology:}  Star formation sources may have dimensions of the order of a few tenths of an arcsecond or a few kiloparsecs at cosmological distances while SMBH driven AGN sources are typically very small, of the order of 0.001 arcsec (10 pc) or less, and are coincident with the galaxy nucleus or QSO. Quasars and AGN powered by SMBHs are often variable on time scales as short as days with corresponding changes in their morphology indicating highly collimated outflows with apparent superluminal velocity. SMBH driven AGN may also contain extended lobes tens or hundreds of kiloparsecs distant from the compact nucleus, and sometimes show optical and radio jets joining the nucleus and radio lobes. 

$\bullet$\ {\bf Radio Spectra:}  Star forming sources and the extended radio lobes of AGN generally have steep radio spectra.  Due to synchrotron self absorption, the compact sources generally have flat or even inverted spectra.

$\bullet$\ {\bf Brightness Temperature:} Star forming sources mostly have measured brightness temperature up to $\sim 10^6$ K while the compact flat spectrum sources associated with AGN have brightness temperature $10^{11-12}$ K or more.

$\bullet$\ {\bf Radio Luminosity:}  Star forming regions typically have a radio luminosity close to $10^{22-23}$ W/Hz and follow the well known correlation between radio and FIR luminosity \cite{C92}.  Radio galaxies and quasars driven by SMBHs may be $10^{4-5}$ times more luminous so their radio luminosity greatly exceeds that expected from the radio/FIR relation characteristic of star forming regions.

$\bullet$\ {\bf X and $\gamma$-ray emission:} Star forming regions are only weak x-ray sources with typical luminosity $\sim 10^{42}$ ergs/sec while SMBH driven AGN can be strong x-ray, $\gamma$-ray, and TeV sources.

$\bullet$\ {\bf Host Galaxies:}    SMBH driven radio sources are located in the nuclei of elliptical galaxies or are associated with quasars which themselves are thought to be the bright nuclei of elliptical galaxies that greatly outshine their host galaxy. Low (optical) luminosity AGN are typically found in early type spiral (often classified as Seyfert) galaxies. Radio emission from star forming regions is typically associated with spiral galaxies but may also be found in the host galaxies of radio quiet quasars (see Section 4).
  
\section{Early Evidence for AGN and SMBHs}

Perhaps the first suggestions that the nuclei of galaxies may contain more than just stars came from Sir James Jeans in 1929 who remarked in his book on Astronomy and Cosmogony \cite[(Jeans 1929)]{2014IAUS..304...78K}, \\

\begin{quote} The centres of the nebulae are of the nature of singular points at which matter is poured into our universe from some other and entirely spatial dimension so that to a denizen of our universe, they appear as points at which matter is being continuously created. \end{quote} \

The modern understanding of the important role of galactic nuclei probably began with the famous paper by Karl Seyfert (1943) who reported on his study of broad strong emission lines in the nucleus of seven spiral nebulae.  Interestingly, although Seyfert's name ultimately became attached to the broad category of spiral galaxies with active nuclei, his 1943 paper received no citations until 1951, and apparently went unnoticed until Baade and Minkowski \cite[(1954)]{BM54} drew attention to the similarity of the Cygnus A radio source spectrum with that of the galaxies studied by Seyfert. 
 
Not until the 1949 Nature paper by Bolton, Stanley, and Slee 
\cite[(1949)]{BSS49} 
did astronomers finally recognize the vast energy requirements of radio galaxies.  Bolton et al. had identified three of the strongest discrete radio sources with the Crab Nebula, M87, and NGC~5128,  Until that time the discrete radio sources were widely thought to be associated with galactic stars.  This was understandable, as Karl Jansky and Grote Reber had observed radio emission from the Milky Way.  The Milky Way is composed of stars, so it was natural to assume that the discrete radio sources had a stellar origin.  Bolton et al.  understood the importance of their identification of the Taurus A radio source with the Crab Nebula which was widely recognized as the remnant of the 1054 supernova reported by Chinese observers. BSS correctly identified two other strong sources with M87 and NGC 5128, but realizing that if they were extragalactic, their absolute radio luminosity would need to be a million times more luminous than that of the Crab Nebula, they  argued that ``NGC 5128 and NGC 4486 (M87) have not been resolved into stars, so there is little direct evidence that they are true galaxies.'' So they concluded that they are within our own Galaxy.   Indeed their paper carried the title ``Positions of Three Discrete Radio Sources of Galactic Radio Frequency Radiation."  John Bolton later argued that he really did understand that M87 and NGC 5128 were very luminous radio sources, but that he was concerned that  that in view of their apparent extraordinary radio luminosity, Nature might not publish their paper.

The following years saw the identification of more radio galaxies, and the changed paradigm which had previously considered all discrete radio sources to be stellar to one with most high latitude sources were assumed to be extragalactic.  The energy requirements were exacerbated in 1951 with the identification of Cygnus A, the second strongest radio source in the sky with a magnitude 18 galaxy at what was then considered a high redshift of 0.056 and a corresponding radio luminosity about $10^3$ times more luminous than M87 and NGC~5128 \cite[(Baade and Minkowski 1954)]{BM54}.  The total energy contained in relativistic particles and magnetic fields in the radio lobes of Cygnus A and other powerful radio galaxies was estimated to be at least $10^{60-61}$ ergs \cite{B59}.  

Hoyle, Fowler, Burbidge and Burbidge \cite[(1964)]{HFBB64} were apparently the first to call attention to gravitational collapse as a possible energy source to power radio galaxies.  By the middle of 1960, many radio sources had been identified with galaxies having red shifts up to 0.24 \cite{B60}.  Typically the optical counterpart of strong radio sources was identified with an elliptical galaxy that was the brightest member of a cluster.  In 1960, Rudolph Minkowski \cite[(1960)]{M60} identified 3C~295 with a mag 20 galaxy at z=0.46.  3C~295 is about ten times smaller than Cygnus A and ten times more distant consistent with the idea that the smallest radio sources might be path finders to finding very distant galaxies. But a few months later Caltech radio astronomers identified the first of several very small sources with what appeared to be galactic stars, thus raising questions about the extragalactic nature of other small diameter radio sources.

\section{The First Quasars}

While searching for ever more distant radio galaxies, Caltech radio astronomers John Bolton and Tom Matthews identified 3C 48 with an apparent stellar object.  At the 107th meting of the American Astronomical Society held in New York in December 1960,  Allan Sandage \cite{S60} reported the discovery of ``The First True Radio Star."    Before he left to return to Australia, John Bolton \cite{B90} speculated that 3C~48 had a high redshift of 0.37, but was apparently dissuaded by Jesse Greenstein and Ira Bowen on the grounds that there was a 3 or 4 Angstrom discrepancy among the corresponding rest wavelengths.  In a  subsequent analysis of the complex emission line spectrum, Jesse Greenstein \cite{G62} interpreted the 3C 48 spectrum in terms of emission lines from highly ionized states of rare earth elements.  He briefly considered a possible redshift of 0.37, but quickly dismissed the possibility that 3C~48 was extragalactic.   Nearly two years would pass, and other compact radio sources would be identified as galactic stars before a series of lunar occultations  would lead to the identification of 3C~273 with a star like object at a redshift of 0.16 and the immediate realization that 3C~48 was also extragalactic with a redshift of 0.37 leading to  the recognition of quasi stellar radio sources or ``quasars''  as the extremely bright nuclei of galaxies. The apparent high radio as well as optical luminosity of quasars, coupled with their very small dimensions presented a further challenge to understanding the source of energy and how this energy is converted to relativistic particles and magnetic fields.

\section{Radio Loud and Radio Quiet Quasars}

The following years led to the identification of more quasars at ever larger redshifts and the suggestion that quasars are powered by accretion onto super massive black holes (SMBH) with masses up to $10^9$ solar masses or more \cite{LB69}. Generally, the identified quasars had a significant UV excess compared with stars, so due to the redshift of their spectrum, they appeared blue on photographic plates facilitating their identification with radio sources with even modest position accuracy.  

   In 1965, Sandage noted that the  density of blue stellar objects on the sky was some thousand time greater than that of 3C radio sources.  Sandage argued that what he called ``quasi stellar galaxies'' are related to quasars, except that they are not strong radio sources.  But, his paper was widely attacked, perhaps in part because of the perceived irregular treatment by the Astrophsyical Journal.  Sandage's paper was received on May 15, 1965 at the Astrophysical Journal, but S. Chandrasekar, the ApJ  editor was apparently so impressed by Sandage's claim for a ``New Constituent of Universe" that he held up publication of the Journal, and Sandage's paper appeared in the May 15 issue.  Tom Kinman \cite{K65} along with Lynds and Villere \cite{LV65} argued that most of Sandage's Blue Stellar Objects were only blue galactic stars, while Fritz Zwicky \cite{Z65} pointed out that he had previously called attention to this phenomena, and he later accused Sandage of ``one of the most astounding feats of plagiarism'' \cite{Z71}.

As it turned out, most of Sandage's Blue Stellar Objects were just that, ``blue stellar objects,"  and only some ten percent of optically identified quasars  are strong radio sources.  But, it has now been more than half a century since we have divided quasars into the two classes of radio loud and radio quiet quasars, and it has still not been clear if there are two distinct populations or rather whether the radio loud population is merely the extreme end of a continuous distribution of radio luminosity. Proponents of each interpretation claim that the other interpretation is due to selection effects.

Many of the previous investigations designed to distinguish between radio loud and radio quiet quasars were limited by contamination from low luminosity AGN with absolute optical magnitudes greater than -23, biased samples based on radio rather than optical selection criteria, and inadequate sensitivity to detect radio emission from most of the radio quiet population.  

In an attempt to overcome these limitations, Kimball et al. \cite{K11} observed 179 quasars selected from the SDSS. All of the these quasars were within the redshift range 0.2 to 0.3 and were brighter than $M_{i}$ = -23 so were genuine quasars that presumably contained a SMBH.  The observations were made with the Jansky Very Large Array at 6 GHz reaching an rms noise of 6 $\mu$Jy. All but about 6 quasars were detected as radio sources with an observed radio luminosity sharply peaked between $10^{22}$ and $10^{23}$ Watts/Hz characteristic of the radio luminosity typically observed from star forming galaxies.  About ten percent of the SDSS sample are strong radio sources with radio luminosities ranging up to $10^{27}$ W/Hz.  Kimball et al.  concluded that the radio emission from radio quiet quasars is due to star formation in the host galaxy.  Similar conclusions were reached by Padovani et al. (2011, 2014) based on the identification and classification of the microJy radio sources found in a deep VLA survey of the Extended Chandra Deep Field South. Based on radio, optical, IR, and X-ray data, Padovani et al. concluded that the microJy radio emission from AGN, like that of galaxies, is  powered primarily by starbursts, and not the SMBHs which powers the AGN. Condon et al.   \cite{C13} argue that these starbursts are fueled by the same gas that flows into the SMBH that powers the quasar and thus accounts for the co-evolution of star formation and SMBHs.

\section{Jet Kinematics and Relativistic Beaming}

Shortly after the recognition of quasars, radio source observations in both the Soviet Union 
\cite{S65}
and the U.S.
\cite{D65}
demonstrated variability on time scales of months or less.  This presented a problem.  Causality arguments suggested linear dimensions, d $\leq$ c$\tau$ where c is the speed of light and $\tau$ the characteristic time scale of the observed variability.  Knowing the quasar redshift and corresponding distance puts a limit to the angular size which for many variable sources was only $\sim 10^{-5}$ arcseconds and the corresponding lower limit to the brightness temperature which appeared to be significantly in excess of the inverse Compton limit of $\sim 10^{11.5}$ K 

For most variable sources, the apparent violation of the inverse Compton limit is now understood in terms of relativistic beaming.  Due to relativistic effects, we observe apparent jet speeds,
luminosities, and brightness temperatures which are related to the
corresponding intrinsic quantities in the AGN rest frame
through the Doppler factor, $\delta$, the Lorentz factor, $\gamma$, and
the jet orientation, $\theta$, with respect to the line of sight \cite{C07}.

The apparent  velocity transverse to the line of sight, $\beta_\mathrm{app}$, the apparent
luminosity, $L$, the apparent brightness temperature, $T_\mathrm{app}$
and the Doppler factor, $\delta$, can be calculated from the Lorentz
factor, $\gamma$,  $\theta$,  and the
intrinsic luminosity, $L_o$.  The apparent transverse velocity $\beta_\mathrm{app}$ is given by

\begin{equation}
\beta_\mathrm{app} = \frac{\beta\sin\theta}{1-\beta\cos\theta}\,, 
\label{eq:beta_app}
\end{equation}
where $\beta = v/c$

For small values of $\theta$, because the radiating source is almost catching up with its own radiation, equation 5.1 shows that the apparent transverse can exceed the speed of light, which is commonly referred to as "superluminal motion."  The apparent luminosity, $L$, is given by
\begin{equation}
L = L_o \delta^n\,,
\label{eq:lum}
\end{equation} 
where the Doppler factor, $\delta$, is  
\begin{equation}
\delta = \gamma^{-1}(1-\beta\cos\theta)^{-1}\,,
\label{eq:delta}
\end{equation}
and where
$L_o$ is the luminosity that would be measured by an observer in the AGN
frame. The quantity $n$ depends on the geometry and spectral index and is
typically in the range between 2 and 3. 

The Lorentz factor, $\gamma$, is  given by 
\begin{equation}
\gamma = {(1-\beta^2)}^{-{1/2}}.
\label{eq:lorentz}
\end{equation}

Quasars or AGN with highly Doppler boosted relativistic jets pointed nearly along the line-of-sight are often referred to as ``blazars.'' Blazars are characterized by rapid flux density variability, apparent superluminal motion, and strong x-ray and $\gamma$-ray emission.  High resolution observations of blazars  provide unique insight to the process by which relativistic jets are accelerated and collimated in the region close to the SMBH.  We want to understand:

$\bullet$ How and where is the relativistic beam accelerated and collimated into narrow jets?  Are there accelerations or decelerations?  Do all parts of the jet move at the same speed?

$\bullet$ What causes the curvature of jets?  Does the flow follow a curved trajectory or is the motion ballistic and characteristic of a rotating nozzle?

$\bullet$ Does the observed apparent velocity reflect the true bulk velocity of motion?   What determines the jet velocity?  Is the velocity related to other properties such as radio, optical, x, or $\gamma$-ray luminosity?  

$\bullet$  What is the maximum observed brightness temperature?  Does it exceed the inverse Compton limit?

$\bullet$ What is the energy production mechanism?

$\bullet$  What can we learn from radio observations about the nature of the SMBH?

\

Very Long Baseline observations made since 1971 have confirmed the apparent superluminal motion expected from highly relativistic bulk motion.  Since 1995, the NRAO Very Long Baseline Array (VLBA) has been used to study the motions of a large sample of quasars and AGN at 7mm by a group from Boston University \cite{M12} and by the international MOJAVE group  \cite{ K04,L09,H09,LAA13,H14}.  More detailed information may be found on the respective web sites:

Boston University 7 mm program:  {\bf http://www.bu.edu/blazars/VLBAproject.html}

MOJAVE 2 cm program: {\bf: //www.physics.purdue.edu/~mlister/MOJAVE/}

\

The results of these programs may be summarized as follows.

\

$\bullet$ Radio loud quasars and AGN show highly relativistic bulk motion with a broad distribution of apparent velocities. In general, the jets appear one sided, probably due to differential Doppler boosting so that the approaching jet appears much brighter than the receding one. Each jet appears to have its own characteristic velocity but there is an appreciable spread in the apparent velocity of the different features within a given jet.  The typical apparent velocity, $\beta_{app} \sim 8$ corresponding to an intrinsic value of $\beta \sim 0.99$.  The maximum observed apparent velocity, $\beta \sim 50$ corresponds to an intrinsic value of $\beta \sim 0.999$.  The parent jet population is mostly only mildly relativistic, but is under represented in  flux density limited samples due to the effect of Doppler boosting.

$\bullet$  The jets with the fasted apparent velocities have the highest apparent luminosity, likely reflecting a correlation between intrinsic speed and intrinsic luminosity rather than simply being the result of Doppler boosting (Cohen et al. 2007).

$\bullet$ Apparent inward motions are uncommon and are likely the result of a feature moving outward along a curved trajectory approaches the line-of-sight so that the apparent separation from the jet base transverse to the line-of-sight appears to decrease with time.

$\bullet$ Individual jet features may show both apparent accelerations and decelerations.  Both the apaprent speed and direction of motion may change with time, but changes in speed are more common than changes in direction, indicating real changes in the Lorentz factor as features propagate down the jet.  In general the apparent speed is greater further down the jet, so that acceleration must take place at distances at least up to $\sim$ 100 pc from the base of the jet (Homan et al. 2009, 2014).

$\bullet$  Many jets show a curved structure and in a few cases there is evidence of an oscillatory behavior.  Sometimes the flow appears to follow pre-existing channels; other times the flow appears ballistic as from a rotating nozzle, perhaps due to precession possibly resulting from a binary black hole pair (Lister et al. 2013).

$\bullet$  In some cases the direction of ejection appears to vary within a well defined cone forming what appears to be an edge brightened jet (Lister et al. 2013) such as shown by the jet in the nearby radio galaxy M87 where there is sufficient linear resolution to resolve the jet transverse to its structure \cite{K07}.

$\bullet$  There appears to be a relation between radio and gamma ray emission.  There is statistical evidence that radio outbursts follow a $\gamma$-ray event by $\sim$ 1 month, but it has been difficult to convincingly establish a one to one correlation between individual radio and $\gamma$-ray events \cite{P10}.

\section{Brightness Temperature Issues}

As described above, inverse Compton scattering limits the maximum observed brightness temperature, T $ < \sim 10^{11.5}$.  At the inverse Compton limit, the energy contained in relativistic particles greatly exceeds that in the magnetic field which is perhaps not unreasonable in a very young source.  If the particle and magnetic energies are in equilibrium, then the corresponding brightness temperature is only $\sim 10^{10.5}$.  

The observed brightness temperature may be calculated from \cite{K06}

\begin{equation} 
 T_b  = \frac{2\ ln 2}{k\pi}\frac{S\lambda^2}{\theta^2} ~K = 1.4\times10^9S\frac{\lambda^2}{\theta^2}(1+z)~K\,,
\label{eq:Tb}
\end{equation}
where k is the Boltzman constant, $\theta$ the angular size in milliarcsec, S is the flux density in Janskys, and $\lambda$ the wavelength in cm.  The resolution of a radio interferometer, is given by the ratio of the observing wavelength, to the interferometer baseline, D; or $\theta$ =  $\lambda$/D.  Putting this back into eqn. 6.1 gives

\begin{equation} 
T_b  = 80SD^2(1+z)~K\,,
\label{eq:Tb}
\end{equation}
so the maximum brightness temperature that may be measured depends only on the flux density and baseline length, and is independent of wavelength.  For ground based observations with a maximum baseline of $\sim 8,000$ km, the highest brightness temperatures which can be reached are $\sim 10^{13}$ K.  Recent observations with the Russian RadioAstron space VLBI satellite have suggested lower limits to brightness temperatures of 3C~273 and other sources $\sim 10^{14}$ K (Kellermann et al. 2014, Kovalev 2014), or  at least two to three orders of magnitude greater than the limit set by inverse Compton cooling.  Several explanations are possible \cite{KIK14}.

1. For a relativistically beamed source the apparent brightness temperature is boosted by a factor $\delta$.  To explain the high observed brightness temperatures in this way would require Doppler factors, $\delta \sim$ $10^2$ to $10^3$.  But typical observed values of $\delta \sim \gamma \sim 10$ with maximum observed values $\sim 50$, and for 3C 273, $\gamma \sim 15$ (Lister et al. 2013). Possibly the bulk flow which is related to the Doppler boosting might be much greater than the pattern flow observed by the VLBA, but Cohen et al. (2007) have shown that this is unlikely.
 
2. The  observed emission may be coherent such as observed in pulsars or the Sun including possible stimulated synchrotron  emission.

3) The radio emission might be the result of synchrotron emission from protons rather than electrons which would enhance the upper limit to the brightness temperature by about the ratio of the proton to electron mass or more than a factor of 1000.  However, proton synchrotron radiation would require a magnetic field strength more than $10^6$ times stronger than needed for electron synchrotron radiation of the same strength at the same wavelength.

4) There may be a continuous acceleration of relativistic particles which balances the energy losses due to inverse Compton cooling.

\section{Summary and Issues}

About ten percent of quasars and bright elliptical galaxies are strong radio sources with radio luminosity, $P_r > 10^{23}~W/Hz$ and are thought to be driven by accretion onto a SMBH.  The weaker radio sources, $P_r < 10^{23}~W/Hz$ are mostly due to star formation in the host galaxy.    The observed properties of radio jets can be interpreted in terms of a highly relativistic outflow from a central engine driven by a SMBH of up to $10^9$ solar masses. But, many questions remain.

$\bullet$ Most quasars and AGN are not strong radio sources.  Why are only $\sim$ 10\% of quasars strong radio sources, although all quasars presumably contain a SMBH to account for their extraordinary optical luminosity.

$\bullet$ How do SMBHs generate relativistic jets?

$\bullet$ How are the jets confined and shaped as they propagate away from the SMBH?  What are the relative roles of velocity shear, hydrodynamic turbulence, shocks, and plasma instabilities in shaping the form and kinematics of relativistic jets?  

$\bullet$  Why do only some jets produce $\gamma$-rays?  How and where are the $\gamma$-rays produced?  What is the relation between radio and $\gamma$-ray emission?  

$\bullet$ Is there evidence for binary black hole pairs?  See the paper by \cite{E15} in this volume.  

$\bullet$ If confirmed, observations of 3.3 GHz sky brightness combined with the density of faint sources detected in deep VLA observations suggest the possible existence of new population of faint radio sources  not due to star formation or to  AGN and unrelated to any known galaxy population?  

$\bullet$  Are there other emission processes which play a role beside incoherent synchrotron radiation?

\section{Acknowledgment}

The National Radio Astronomy Observatory is operated by Associated Universities, Inc. under cooperative agreement with the National Science Foundation.  I am indebted to many colleagues, especially Ron Ekers and members of the MOJAVE team for numerous discussions that have contributed to this paper. 
\\\\

\end{document}